\documentclass[journal]{IEEEtran}

\usepackage{amsmath,graphicx,mathrsfs}
\usepackage{etex}
\usepackage{float,subfigure}
\usepackage{cite,hyperref}

\usepackage{cite}
\usepackage{amsmath,amssymb,amsfonts}
\usepackage{algorithmic}
\usepackage{graphicx}
\usepackage{textcomp}
\usepackage{booktabs}
\usepackage{multirow}
\usepackage{footnote}
\usepackage{stfloats}

\newcommand {\ba}[1]{\raisebox{-1.5ex}[0pt][0pt]{\shortstack{#1}}}

\ifCLASSINFOpdf
\else
\fi

\begin{document}
%
\title{Advanced Geometry Surface Coding for Dynamic Point Cloud Compression}

\author{Jian~Xiong,~\IEEEmembership{Member,~IEEE},
~Hao~Gao,~\IEEEmembership{Member,~IEEE},~Miaohui~Wang,~\IEEEmembership{Member,~IEEE},~Hongliang~Li,~\IEEEmembership{Senior~Member,~IEEE},~King~Ngi~Ngan,~\IEEEmembership{Fellow,~IEEE}~and~Weisi~Lin,~\IEEEmembership{Fellow,~IEEE}
\thanks{Manuscript received 12. 24, 2020. This work was supported in part by the National Natural Science Foundation of China (No. 61701258 and No. 61931012).~(\emph{Corresponding authors: Hao~Gao and Miaohui~Wang}.)}
\thanks{Jian~Xiong is with College of Telecommunications and Information Engineering, Nanjing University of Posts and Telecommunications, Nanjing, China, 210003, (e-mail: jxiong@njupt.edu.cn).}
\thanks{Hao~Gao is with College of Artificial Intelligence, Nanjing University of Posts and Telecommunications, Nanjing, China, 210003, (e-mail: tsgaohao@gmail.com).}
\thanks{Miaohui~Wang is with School of Computer Science and Cyberspace Security, Hainan University, and is also with Guangdong Key Laboratory of Intelligent Information Processing, Shenzhen University (E-mail: wang.miaohui@gmail.com).}
\thanks{Hongliang~Li and KingNgi Ngan are with School of Information and Communication Engineering, University of Electronic Science and Technology of China,
 Chengdu, China, 611731, (e-mail: hlli@uestc.edu.cn and knngan@ee.cuhk.edu.hk).}
 \thanks{Weisi~Lin is with School of Computer Science and Engineering, Nanyang Technological University, Singapore, 639798 (e-mail: wslin@ntu.edu.sg).}
}

\maketitle

\begin{abstract}
In video-based dynamic point cloud compression (V-PCC), 3D point clouds are projected onto 2D images for compressing with the existing video codecs. However, the existing video codecs are originally designed for natural visual signals, and it fails to account for the characteristics of point clouds. Thus, there are still problems in the compression of geometry information generated from the point clouds. Firstly, the distortion model in the existing rate-distortion optimization (RDO) is not consistent with the geometry quality assessment metrics. Secondly, the prediction methods in video codecs fail to account for the fact that the highest depth values of a far layer is greater than or equal to the corresponding lowest depth values of a near layer. This paper proposes an advanced geometry surface coding (AGSC) method for dynamic point clouds (DPC) compression.
The proposed method consists of two modules, including an error projection model-based (EPM-based) RDO and an occupancy map-based (OM-based) merge prediction. Firstly, the EPM model is proposed to describe the relationship between the distortion model in the existing video codec and the geometry quality metric. Secondly, the EPM-based RDO method is presented to project the existing distortion model on the plane normal and is simplified to estimate the average normal vectors of coding units (CUs).
Finally, we propose the OM-based merge prediction approach, in which the prediction pixels of merge modes are refined based on the occupancy map. Experiments tested on the standard point clouds show that the proposed method achieves an average 9.84\% bitrate saving for geometry compression.
\end{abstract}

\section{Introduction}

\IEEEPARstart{A}{}point cloud is defined as a set of 3D points, in which each point is expressed as a 3D coordinate and specific attributes (such as texture and normal). With the advancement of 3D capture technology \cite{usecase}, point clouds are widely used in the applications, such as virtual reality, immersive telepresence\cite{Fuchs2014Comp}, mobile localization\cite{ChengTIP2017,ChengTIP2019}, and 3D printing. A typical use of point clouds is to represent holographic images of humans in virtual reality and immersive telepresence \cite{JHou2020TIP,HuiYuan_2019TB}. We call it dynamic point clouds (DPC)\cite{ZhanMa2020TCSVT,YXu2020TCSVT}.
These applications with DPC generate a huge amount of data. For an uncompressed DPC with one million points in each frame, its bitrate will reach up to 180MB/s \cite{Lili2020TIP} if its frame rate is 30fps. Therefore, compression of DPC becomes a critical part of these emerging 3D systems.

Since a DPC represents a continuously moving object, consecutive frames in the DPC generally have strong temporal redundancy.
Recent works have been concerned with motion estimation (ME) and motion compensation (MC) to reduce temporal redundancy.
Specifically, ME and MC are performed on either 3D cubes \cite{Thanou2016TIP,TIP_2017_MCC,Mekuria_2017CSVT,Dorea2019ICIP} or 2D blocks  \cite{lasserre2017technicolor,budagavi2017samsungs,schwarz2017nokias,He_2017_APCC}.
However, the point clouds are with irregular partitions such that some points in consecutive static point cloud (SPC) frames may not have explicit correspondences.
Therefore, the 3D ME-based methods cannot fully exploit temporal correlations.
For better keeping the temporal correlation, 2D ME-based methods \cite{lasserre2017technicolor,budagavi2017samsungs,schwarz2017nokias,He_2017_APCC} try to project the 3D point clouds onto 2D space such as cylinders and cube faces, and then organize the projected samples into 2D videos for performing video compression. However, these methods may lose a large number of points due to occlusion in projection.

To improve the ability to maintain the temporal correlation and increase the number of projected points,
a patch projection method was proposed to convert the DPC into 2D videos\cite{mammou2017video}.
Specifically, the input DPC is decomposed into multiple patches according to the similarity of normals.
These patches are packed into 2D grids to generate a geometry video and an attribute video, respectively.
Since the patches are with irregular shapes, there are empty samples in the 2D grids, hence occupancy maps are generated to indicate whether the samples belong to the patches or not\cite{vZak}.
Then, the occupancy maps and the generated videos are compressed using existing video codecs, such as High-Efficiency Video Coding (HEVC) \cite{Sullivan_CSVT2012}.
This method is called a video-based dynamic point cloud compression (V-PCC)\cite{jang2019video}.
Since it can achieve a better trade-off between the ability to maintain the temporal correlation and the number of projected points, it is the winner of the MPEG call for proposals of DPC compression \cite{preda2017report}, and then is integrated into the MPEG V-PCC reference software TMC2\cite{TMuC2}.

In V-PCC, to handle the case that multiple points are projected on the same sample, each patch is projected onto two images, i.e., the corresponding near and far layers. The near layer stores the lowest depth and the far layer stores the highest depth \cite{Schwarz2018JETCAS}.
It is known that the points with the highest depth values form the surface of a point cloud. Thus, we call them as \emph{surface points}.
Points in point clouds often represent the surface and point clouds are characterized by surfaces of structures.
High encoding performance of the surface points is crucial for the quality of geometry information.
The geometry information is evaluated by using two metrics, including a point-to-point error (D1) \cite{EvaCriteria} and a point-to-plane error (D2) \cite{D2error}.
Because of this, the D2 error metric which accounts for the surface structures can better track subjective visual quality than the D1 error metric \cite{D2error}.


Based on V-PCC, recent works \cite{Lili@TMM2020,JLiuICME2019,Lili2020TCSVT,Lili2020TIP,OMGVPCC2021TCSVT,Lili@TMM2020rate} were proposed to be applied on its sub-modules, such as patch packing, motion estimation, rate-distortion optimization (RDO), and rate control.
These methods improve compression performance further.
However, the existing video codec used in V-PCC is originally designed for natural visual signals, and it fails to account for the characteristics of point clouds.
Especially, there are still some problems in coding the surface points.
Firstly, the distortion model in the existing RDO is not consistent with the quality metric of geometry information.
In the existing RDO, the distortion is computed based on the distance between the reconstructed signal and its original signal.
However, the objective quality of geometry is measured based on the distance between the reconstructed point and its corresponding point (the nearest neighbor) in the reference point cloud (as the D1 and D2 metrics), while the corresponding point may not necessarily be the original point.
Secondly, in V-PCC, the near layers are used as the reference frames of the corresponding far layers.
Especially for the skip or merge mode with a zero motion vector (MV), the reconstructed lowest depth values are used as the prediction of the corresponding highest depth values. However, the highest depth value of each sample is always greater than or equal to the lowest depth value.
The existing prediction coding methods do not take this characteristic into account.

\begin{figure*}[tpb]
  \centering
  \centerline{\includegraphics[width=14cm]{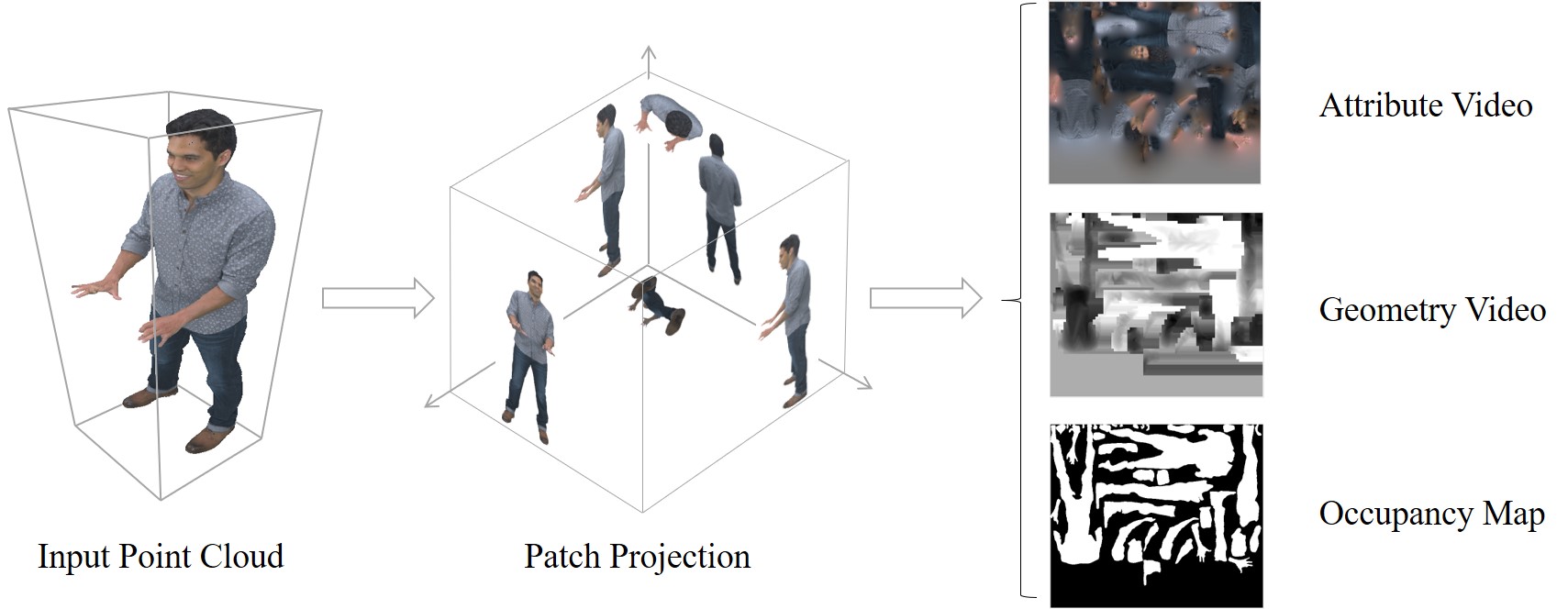}}
\caption{A diagram of the patch-projection-based DPC compression}\centering
\label{fig:vpcc}
\end{figure*}

Therefore, in this paper, we propose an advanced geometry surface coding (AGSC) method to solve the aforementioned two problems. The proposed method is composed of two modules, including an error projection model based RDO algorithm and an occupancy map-based merge prediction approach.
The main contributions are listed as follows.
\begin{enumerate}
\item
We identify a problem that the existing distortion model is not consistent with the geometry quality metrics.
We present an error projection model (EPM) to describe the relationship between the distortion model in the existing video codec and the geometry quality metric. In this model, by approximating a coding unit (CU) as a plane, the D2 error is computed as projecting the existing distortion model on the plane normal.
\item
We present an EPM-based RDO for improving geometry DPC compression.
The proposed EPM-based RDO method is simplified to estimate the average normal vector of CUs.
Then, the least-square estimation is applied to estimate the normal by calculating the gradients.
\item
We propose an occupancy-map (OM)-based merge prediction approach, in which the prediction pixels are refined based on the occupancy map.
This approach is designed based on the fact that the depth values of the far layers are greater than or equal to that of the corresponding near layers.
An analysis shows that a refined prediction error is smaller than the original prediction error.
\end{enumerate}
Experiments are performed on typical dynamic point clouds. The results show that the proposed AGSC method can significantly improve the compression of geometry information.
Furthermore, ablation experiments demonstrate that both of the two modules are efficient for improving the coding performance.

The rest of the paper is organized as follows. In Section II, some related works are reviewed.
Then, the background of V-PCC is introduced in Section III.
In Section IV, the EPM mode is proposed by investigating the existing distortion model and the geometry quality metrics.
In Section V, we introduce the EPM-based RDO method and its implementation.
The improved merge prediction methods are proposed in Section VI.
Experimental results are provided in Section VII to verify the proposed method.
Finally, we make a conclusion in Section VIII.


%

\section{Related Work}
\label{sec:rw}

\subsubsection{DPC Compression with 3D ME}

To fully reduce the temporal redundancy of DPC, recent works mainly try to perform ME in either 3D or 2D spaces.
For the 3D-based ME, Kammerl et al. \cite{Kammer_ICRA2012} proposed a method for decomposing the point clouds into octree data structures and then encoding the differences of octree structure of consecutive point clouds.
Daribo et al. \cite{daribo2012efficient} presented a point cloud encoder based on a curve-based representation scheme. The curve-based intra and inter prediction modes were utilized to remove the spatial and temporal correlation of point clouds. However, the point cloud frames have varying numbers of points, such that various points have no explicit correspondence in the neighboring SPCs.
In \cite{Thanou2016TIP}, a DPC is represented as a set of graphs, in which the nodes denote the 3D positions and color attributes. Then the 3D ME is formulated as a feature matching problem between successive graphs. However, the estimated MV is inaccurate because some nodes do not have explicit features. Moreover, the feature matching step is computationally expensive.
In \cite{TIP_2017_MCC}, a DPC is decomposed into blocks of voxels, and each block is encoded in either intra or inter prediction mode. Compared with only using intra prediction mode, the simple translational motion estimation can reduce the coding rate to some extent. However, it is still not efficient for representing the 3D rotation in DPC. In \cite{Mekuria_2017CSVT}, the temporal correlations are exploited by registering the neighboring frames via an iterative closest point (ICP) based method.
However, complicated regions are still difficult to be registered.
In summary, the 3D ME methods still cannot fully exploit the temporal correlations of DPCs, due to the varying numbers of points and inflexible partition sizes.

\subsubsection{DPC Compression with 2D ME}
Considering the flexible partition sizes and mature ME in 2D video compression frameworks, some works \cite{lasserre2017technicolor,budagavi2017samsungs,schwarz2017nokias,He_2017_APCC} try to project the 3D point clouds onto 2D videos and then perform 2D ME on the videos.
In \cite{lasserre2017technicolor}, the DPCs are projected into 2D cube faces via an octree-based method.
In \cite{budagavi2017samsungs}, the 3D points are directly sorted and padded into 2D videos.
However, these projection methods will lead to shape changes, such that the temporal correlations are unable to be fully exploited.
In \cite{schwarz2017nokias,He_2017_APCC}, the 3D point clouds are projected onto 2D space as cube faces or cylinders and then organize the projected samples into 2D videos for encoding with a video compression framework. In these methods, the temporal correlation is highly kept but many points may be lost due to occlusion.

\subsubsection{V-PCC and Its Improvements}

To solve the above problems, a patch projection-based method \cite{mammou2017video} was proposed to decompose the DPC into multiple patches according to the similarity of normals. These patches are then organized into 2D videos. The occupancy maps, as well as auxiliary information, are encoded to indicate the pixel and patch locations for reconstruction. This scheme can achieve a good trade-off between the ability to keep the temporal correlation and the number of projected points, such that it wins in MPEG call for proposals for the DPC compression.

On the basis of the V-PCC method, many works were proposed to improve compression performance.
A data-adaptive packing method was proposed by dividing the input sequence into groups adaptively\cite{JLiuICME2019}.
Considering that the projection may destroy the 3D object motion, Li et al. \cite{lili2019cfp,Lili2020TIP} proposed an advanced 3D motion prediction method by using the auxiliary information.
The method can significantly improve the accuracy of motion prediction, such that it has been adopted by the V-PCC encoder.
Since the coding efficiency is reduced by the unoccupied pixels among the patches,
Li et al. \cite{Lili2020TCSVT} proposed an occupancy map based RDO method, in which only the rate is not considered for the unoccupied pixels during the RDO process.
An efficient padding method was proposed in \cite{Lili@TMM2020}. In this method, some of the unoccupied pixels are padded with the real points in the original DPC, such that the number of projected points is increased.
Xiong et al. \cite{OMGVPCC2021TCSVT} proposed an occupancy map guided fast V-PCC coding method, in which a fast CU decision scheme and a fast mode decision scheme are proposed based on the occupancy map.

\section{Background of V-PCC}

In the MPEG V-PCC reference software, the patch-projection-based method decomposes a DPC into a geometry video, an attribute video, and occupancy maps.
As shown in Fig. 1, the adjacent 3D points are divided into patches.
Each patch consists of the adjacent 3D points that are with similar normal vectors.
Then, the attribute and geometry components of these patches are packed on 2D frames, respectively.
Since the patches are within irregular regions, the occupancy maps are generated to indicate which samples are the non-empty points.
Moreover, the packed 2D frames are filled and smoothed with dilation methods.

\begin{figure}[tpb]
  \centering
  \centerline{\includegraphics[width=6cm]{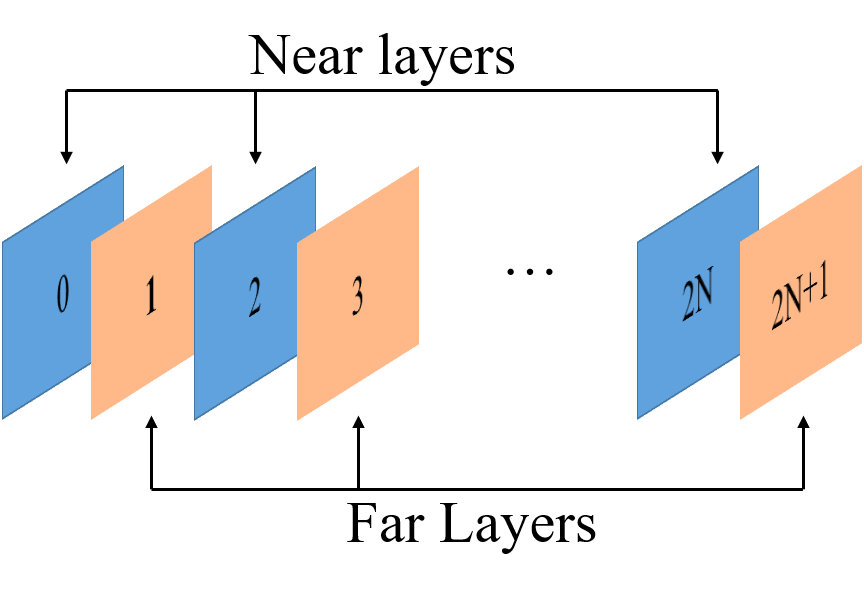}}
\caption{A diagram of the corresponding far and near layers}\centering
\label{fig:farnear}
\end{figure}

Note that multiple points may be projected to the same sample. To handle that, each patch is projected onto two images, i.e., the corresponding far and near layers.
The far and near layers are alternatively organized into 2D videos.
As shown in Fig. \ref{fig:farnear}, the even layer, also called the near layer, stores the lowest depth values.
And the odd layer, i.e., the far layer, stores the highest depth values. Besides, additional conditions are defined to ensure that the differences between the corresponding two layers do not exceed the surface thickness.

Finally, the generated occupancy maps, the geometry video, and the attribute video are encoded with the existing video codec, such as HEVC.

\section{Error Projection Model}
\label{sec:ana}

In this section, we propose the inconsistency between the distortion model in the existing RDO and the geometry quality metrics.
Then, an error projection model is presented to describe the relationship between the distortion model in the existing RDO and the D2 quality metric.

\subsection{Existing Distortion Model in RDO}
\label{ssec:pc2bc}

It is known that many coding modes and flexible CU structures are provided in existing video codec.
For example, HEVC provides many prediction modes, such as intra prediction, skip mode, merge prediction, and the other inter prediction modes.
RDO in the existing video codec is a critical part of choosing the optimal coding mode and CU structure\cite{huanqiangzengCSVT2011,jxiongmrf,lshen2014TIP,yunzhang2015,jxiongTMM2015}.
Specifically, a coding mode with the minimal rate-distortion (R-D) cost is selected as the optimal one. The R-D cost is expressed as,
\begin{equation}
\label{equ:0}
J = D + \lambda R,
\end{equation}
where $J$ denotes the R-D cost, the term $D$ denotes the
reconstruction distortion and $R$ is the number of coding bits.

The reconstruction distortion is calculated as summation of square errors (SSE) between the reconstructed signals and their original signals.
\begin{equation}
\label{equ:1}
D = \sum_{x_k\in \textbf{X}} (x_k- \hat{x_k})^2,
\end{equation}
where $x_k$ and $\hat{x_k}$ denote the original value and reconstructed value of a coding unit (CU) $\textbf{X}$, respectively.

PSNR is widely used for the evaluation of video quality and is computed based on the mean of square errors (MSE) between the reconstructed signals and the original signals.
Thus, considering the PSNR metric, the distortion model in the existing RDO is applicable for video coding.

\subsection{Geometry Quality Metrics}

The quality of geometry information is evaluated by metrics, including PSNR of a point-to-point error (D1 error) \cite{EvaCriteria} and a point-to-plane error (D2 error)\cite{D2error}.
These metrics are computed based on the symmetric hausdorff distance between the reconstructed points and their corresponding points in the reference point cloud.
Note that, the \emph{corresponding point} denotes the nearest neighbor in the reference point cloud.

Specifically, for each point $a_j$ in a point cloud $\textbf{A}$, the corresponding point in the reference point cloud $\textbf{B}$ is denoted as $b_j$,
The point-to-point error is computed based on the distance between $a_j$ and $b_j$, i.e.,
\begin{equation}
\label{equ:2}
e_{\textbf{A},\textbf{B}}^{c2c} = \frac{1}{N_{\textbf{A}}} \sum_{\forall a_j \in \textbf{A}} ||\overrightarrow{b_ja_j}||_2^2,
\end{equation}
where $N_\textbf{A}$ denotes the number of points in \textbf{A}.

To account for the surface structure, the point-to-plane error is computed based on the distance from the point to its reference plane. It is expressed as,
\begin{equation}
\label{equ:3}
e_{\textbf{A},\textbf{B}}^{c2p} = \frac{1}{N_{\textbf{A}}} \sum_{\forall a_j \in \textbf{A}} (\overrightarrow{b_ja_j} \cdot \overrightarrow{N_j})^2,
\end{equation}
where $\overrightarrow{N_j}$ denotes the unit normal vector of $b_j$.

\subsection{Inconsistency between the existing distortion model and geometry quality metrics}

\begin{figure}[tpb]
  \centering
  \centerline{\includegraphics[width=8cm]{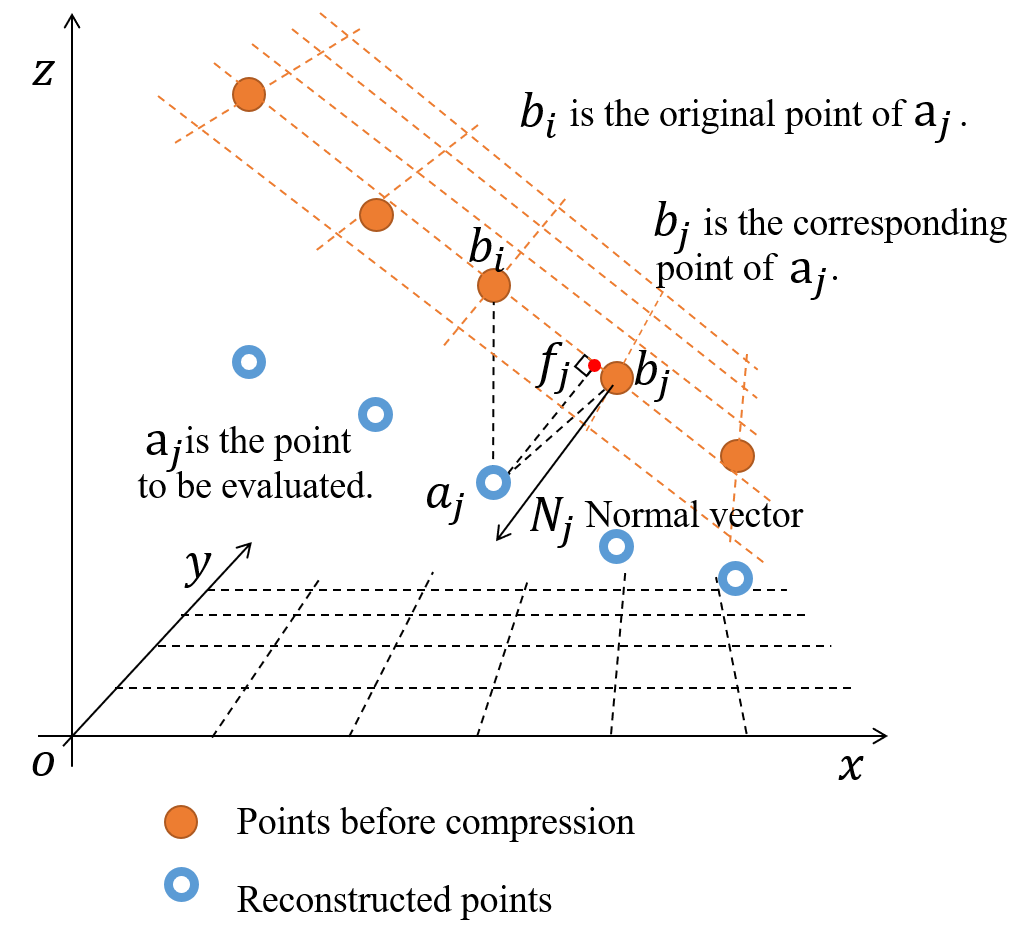}}
\caption{The relationship between the distortion in existing rate distortion optimization and the geometry quality metrics.}\centering
\label{fig:distortion}
\end{figure}

In this work, we consider the coding of the geometry CUs.
As shown in Fig. 3, the solid points denote the points before compression (i.e., original points), and the circle shape points denote the reconstructed points.
For a patch in an original point cloud $\textbf{B}$, when it is compressed with the existing video codec, it becomes a reconstructed patch in $\textbf{A}$.
We also consider a given geometry CU $\textbf{X}$ which belongs to point cloud $\textbf{B}$, i.e., $\textbf{X}\subset\textbf{B}$.
A point $b_i \in \textbf{X}$ is compressed to be a point $a_j$. The distortion in existing RDO is computed based on the distance between $b_i$ and $a_j$, i.e.,
\begin{equation}
\label{equ:4}
D = \sum_{\forall b_i\in \textbf{X}} ||\overrightarrow{b_ia_j}||_2^2.
\end{equation}

Moreover, the corresponding point of $a_j$ is point $b_j$ (not $b_i$), and $N_j$ is the normal vector of $b_j$. Thus, the D1 error is evaluated based on the distance between $b_j$ and $a_j$, i.e., $||\overrightarrow{b_ja_j}||_2$.
Note that, even though the geometry quality metrics are based on the symmetric hausdorff distance, we only consider the case of $\textbf{A}$ to $\textbf{B}$ for simplicity.

Furthermore, the geometry patches describe the surfaces of point clouds.
A local block in a patch is reasonable to be approximated as a plane, i.e., the CU $\textbf{X}$ can also be called as plane $\textbf{X}$.
That is, all the points in $\textbf{X}$ have the same normal vector. The symbol $f_j$ is the foot point of $a_j$ on a plane $\textbf{X}$, and is denoted as the small red point.
That is, $\overrightarrow{N_j}$ and $\overrightarrow{f_ja_j}$ have the same direction.
Thus, the D2 error is computed based on the distance between $f_j$ and $a_j$, i.e.,
$||\overrightarrow{f_ja_j}||_2$.

Note that the foot point is the point in plane $\textbf{X}$ which has the nearest distance to $a_j$,
and it may be not a real point in the reference point cloud.
Thus, it is reasonable that the corresponding point $b_j$ is close to the foot point.
Furthermore, the geometry patch describes the depth values in the orientation of patch projection. In this case, the patch projection plane is expressed as plane $xoy$.
However, for plane $\textbf{X}$ with a large angle to the projection plane, the corresponding point $b_j$ is not necessarily the same as the original point $b_i$ or the foot point $f_j$, i.e.,
\begin{equation}
\label{equ:41}
||\overrightarrow{b_ia_j}||_2 \neq ||\overrightarrow{b_ja_j}||_2~~~~\text{and}~~~~||\overrightarrow{b_ia_j}||_2 \neq ||\overrightarrow{f_ja_j}||_2
\end{equation}
Therefore, the distortion model in the existing RDO is not consistent with the quality metrics (both the D1 and D2 errors) of geometry information.

\subsection{Error Projection Model}

To resolve the inconsistency between the existing distortion model and the geometry quality metrics, this work proposes an error projection model (EPM). To be more specific, since the D2 metric can better track subjective visual quality than the D1 error metric \cite{D2error}, the D2 metric is adopted in this study.

Based on the analysis above, for the coded CU $\textbf{X}$, the D2 error is calculated as,
\begin{equation}
\label{equ:5}
D_{c2p} = \sum_{\forall b_i \in \textbf{X}} ||\overrightarrow{f_j a_j}||_2^2,
\end{equation}
where $D_{c2p}$ denotes the D2 error.
Since $f_j$ is the foot point of $a_j$ on plane $\textbf{X}$.
The distortion $D_{c2p}$ can be rewritten as,
\begin{equation}
\label{equ:6}
D_{c2p} = \sum_{\forall b_i \in \textbf{X}} (||\overrightarrow{b_i a_j}||_2 \cos \angle b_ia_jf_j)^2,
\end{equation}
where $\angle b_ia_jf_j$ denotes the angle between $\overrightarrow{b_i a_j}$ and $\overrightarrow{f_j a_j}$.

It is known that a geometry patch represents the depth values in the patch projection orientation. Thus, the vector from the original point to the reconstructed point is parallel with the projection orientation. In this case, $\forall b_i \in \textbf{X}$, $\overrightarrow{b_i a_j} = (0,0,-1)$.
Moreover, since the local patch can be approximated as a plane, the normal vectors for all the points have the same direction.
That is, for all the points in CU $\textbf{X}$, the angles between their normal vectors and the projection orientation have the same value, which is denoted as $\theta$, i.e., $\angle b_ia_jf_j = \theta, \forall b_i \in \textbf{X}.$
Combining with (\ref{equ:4}) and (\ref{equ:6}), for the CU $\textbf{X}$, we have,
\begin{equation}
\label{equ:7}
D_{c2p} = D \cos ^2 \theta.
\end{equation}
We call it as the error projection model (EPM). In this model, by approximating a CU as a plane, the D2 error is computed as projecting the existing distortion on the plane normal.

\newcounter{TempEqCnt} 
\setcounter{TempEqCnt}{\value{equation}} 
\setcounter{equation}{16} 
\begin{figure*}[hb] 
	\hrulefill  
	\begin{equation}
\left[ \begin{array}{c}
U\\
V\\
W
\end{array}
\right ]
=
\frac{1}{40}
\left[ \begin{array}{cccccccccccccccc}
 -3 & -1 & 1 & 3 & -3 & -1 & 1 & 3 & -3 & -1 & 1 & 3 & -3 & -1 & 1 & 3 \\
-3 & -3 & -3 & -3 & -1 &-1 &-1 &-1 &  1 & 1 & 1 & 1 & 3&3&3&3\\
17.5  & 12.5 & 7.5 & 2.5 & 12.5 & 7.5 & 2.5 & -2.5 & 7.5 & 2.5 & -2.5 & -7.5& 2.5& -2.5&-7.5&-12.5
\end{array}
\right ]\textbf{z}
\end{equation}
\end{figure*}

\section{Error Projection Model-based RDO}

\subsection{Strategies}

Since the distortion model in the existing RDO cannot accurately estimate the quality of geometry information, coding modes generated by RDO cannot achieve the optimal performance.
Therefore, in this work, an EPM-based RDO method is proposed for coding the geometry video.
To be more specific, the D2 metric is directly estimated in the proposed RDO method.
That is, the new R-D cost is calculated based on D2 error, i.e.,
\begin{equation}
\setcounter{equation}{10}
\label{equ:8}
J_{c2p} = D_{c2p} + \lambda R,
\end{equation}
where $J_{c2p}$ is the improved R-D cost.

The proposed EPM model describes the relationship between the existing distortion model and the D2 quality metric as (\ref{equ:7}). For each input CU, the projection orientation and its normal vector is fixed, such that $\theta$ is a constant.  Combining (\ref{equ:7}) with (10), we can have,
\begin{equation}
\label{equ:11}
J_{c2p} = D + \frac{\lambda}{\cos^2\theta} R.
\end{equation}
That is, the EPM-based RDO can be converted to refining the lambda parameter in existing RDO.
Here, a small angle $\theta$ means that the reconstruction distortion is close to the D2 error, i.e., only a subtle change on $\lambda$ is needed.
In contrast, a large angle means a large difference between the reconstruction distortion and the D2 error, i.e., a large change on $\lambda$ is necessary.

Note that, in the existing codec, RDO is to select the optimal CU quad-tree structure and the optimal encoding modes.
It traverses from the root to the nodes. Thus, CUs in the same quad-tree share the same angle $\theta$.
That is, we only need to estimate the normal vectors of CUs in depth 0.

\subsection{Implementation}

The next task is to estimate the angle $\theta$.
In this work, we might as well suppose that the projection orientation is parallel with the vector (0, 0, -1). The task can be converted to estimate the normal vector of each CU.
Each CU is approximated as a plane, denoted as $z = Ux + Vy + W$.
Thus, for all the points in a CU, the normal vectors have the same direction, denoted as $(U, V, -1)$.
The estimation of the normal vector is converted to fitting the plane equation based on all the points of the CU.
The least-square estimation is applied to fit the plane equation.

Let us consider a CU with $n$ points, denoted as $b_i = (x_i, y_i, z_i), i = 0, 1, ..., n-1.$ The method is to minimize the error as follow:
\begin{equation}
\label{equ:9}
min \{L(U, V, W) = \sum_{i=0}^{n-1} (Ux_i + Vy_i +W - z_i)^2\}.
\end{equation}

By taking the partial derivative of $U$, $V$, and $W$ equal to zero, we can obtain,
\begin{equation}
\label{equ:12}
\left[ \begin{array}{c}
U\\
V\\
W
\end{array}
\right ]
=
(\textbf{P}_{xy}\textbf{P}_{xy}^T)^{-1}\textbf{P}_{xy}\textbf{z},
\end{equation}
where the matrix $\textbf{P}_{xy}$ is composed of the coordinates $x_i$ and $y_i$, and the vector $\textbf{z}$ consist of the coordinates $z_i$. They are expressed as,
\begin{equation}
\label{equ:10}
\textbf{P}_{xy}
=
\left[ \begin{array}{cccc}
 x_0 & x_1 & ... & x_{n-1}\\
 y_0 & y_1 & ... & y_{n-1}\\
1 & 1 & ... & 1
\end{array}
\right ],
\end{equation}
and
\begin{equation}
\label{equ:11}
\textbf{z}
=
\left[ \begin{array}{c}
 z_0 \\
 z_1 \\
 ... \\
 z_{n-1}
\end{array}
\right ].
\end{equation}

Based on (\ref{equ:12}), directly estimating the normal vector for the entire $64\times64$ blocks would introduce floating point operations.
For simplicity, a CU with the size $64\times64$ is split into 256 non-overlapped $4\times 4$ blocks.
A normal vector is estimated for each block, and the average vector for these blocks is considered as the normal vector of the CU.

For the points of each $4\times 4$ block, the coordinates $x_i$ and $y_i$ can be considered as the column number and row number. That is, the matrix $\textbf{P}_{xy}$ is a constant matrix, as
\begin{equation}
\label{equ:13}
\textbf{P}_{xy}
=
\left[ \begin{array}{cccccccccc}
 1 & 2 & 3 & 4 & 1&2&3&4& ... & 4\\
 1 & 1 & 1 & 1 & 2&2&2&2&... & 4\\
1 & 1 & 1 & 1 & 1 &1&1&1& ... & 1
\end{array}
\right ].
\end{equation}

Then, the plane parameters $U$, $V$, and $W$ can be obtained as shown in (17). That is, the normal vector $(U, V, -1)$ is estimated by calculating the gradients in X and Y directions with the following two filters, respectively, i.e.,
\begin{equation}
\setcounter{equation}{18}
\label{equ:18}
\frac{1}{40}
\left[ \begin{array}{cccc}
 -3 & -1 & 1 & 3 \\
 -3 & -1 & 1 & 3 \\
 -3 & -1 & 1 & 3 \\
  -3 & -1 & 1 & 3
\end{array}
\right ]~~
\text{and}~~
\frac{1}{40}
\left[ \begin{array}{cccc}
 -3 & -3 & -3 & -3 \\
 -1 & -1 & -1 & -1 \\
 1 & 1 & 1 & 1 \\
  3 & 3 & 3 & 3
\end{array}
\right ].
\end{equation}
It can be seen that only integers are in these filters, such that the gradient calculation is simplified.
Finally, we can obtain the average normal vector denoted as $(\hat{U}, \hat{V}, -1)$, and $\cos \theta = \frac{1}{\sqrt{\hat{U}^2+\hat{V}^2+1}}$. To prevent the estimated angle from becoming too large, we clip $\frac{1}{\cos^2 \theta}$ with a number of 2.

Note that, since the D2 error is designed to measure the point-to-plane distortion, the key technology of the proposed EPM-based RDO is to estimate the normal vector of the surface, which is mainly affected by the surface points.
In this work, we only account for the surface points. That is, the proposed EPM-based RDO method is applied only on the far layers.

\section{Occupancy Map based Merge Prediction}
\label{sec:ana}

%
%
%
%

In this section, we first investigate the inter prediction error with the zero motion vector (MV).
An analysis shows that a refined prediction error is smaller than the original one.
Then, we propose two alternative improved merge prediction approaches.

\begin{figure}[tpb]
  \centering
  \centerline{\includegraphics[width=7cm]{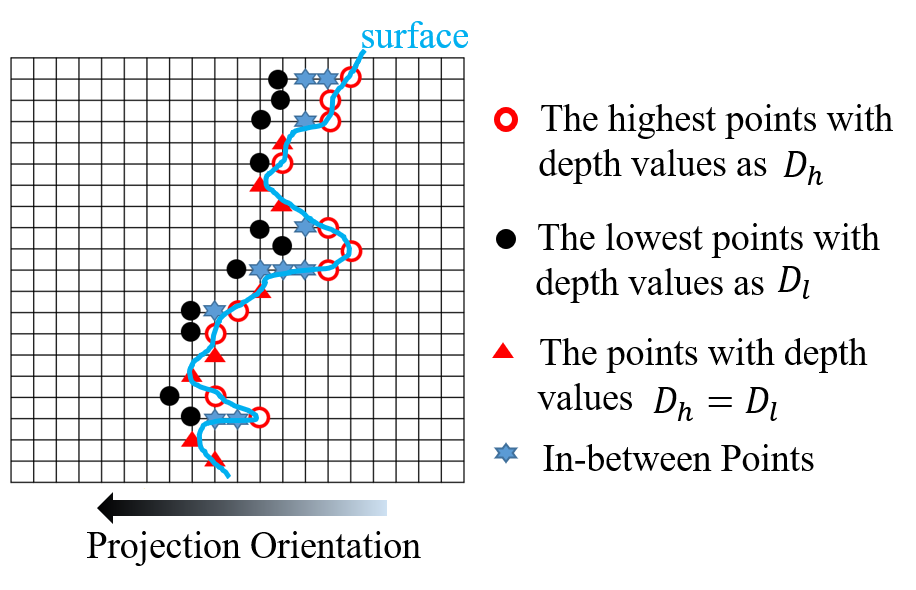}}
\caption{A diagram of the points being projected on the corresponding far and near layers \cite{vZak}.}\centering
\label{fig:points}
\end{figure}

\subsection{Prediction Error of the Far Layers}

In MPEG V-PCC, to handle that multiple points may be projected on the same sample, each patch is projected onto two layers, i.e., the corresponding far and near layers. The near layer stores the lowest depth values denoted as $D_l$, and the corresponding far layer stores the highest depth values denoted as $D_h$.
Note that, a surface thickness $\tau$ is predefined to ensure that $D_h$ is within $[D_l, D_l + \tau]$, i.e., $D_h \geq D_l$ \cite{Schwarz2018JETCAS}.
A diagram of the projected points can be seen in Fig. \ref{fig:points} \cite{vZak}. The circle shape points are those with the highest depth values. The solid points are those with the lowest depth values. Moreover, the star points are the in-between ones.
Whereas the triangle shape points denote the samples projected by only one point.

In V-PCC, the reconstructed near layer is used as a reference frame for coding the corresponding far layer.
Moreover, it is known that a zero motion vector (MV) is generally used as one of the candidate MVs in the skip and merge modes.
Thus, when considering the skip and merge modes with a zero MV, for each sample, the reconstructed lowest depth value is used as the prediction of the corresponding highest depth value.

To be more specific, we denote a CU in a far layer as $\textbf{X}_h$ and the co-located CU in the corresponding near layer as $\textbf{X}_l$.
That is, for each sample $i$, $x_{h,i} \geq x_{l,i}$, where $x_{h,i} \in \textbf{X}_h$ and $x_{l,i} \in \textbf{X}_l$.
We denote the coding error of $x_{l,i}$ as $n_{l,i}$, $n_{l,i} \in \textbf{N}_l$.
That is, the reconstructed lowest depth value is $x_{l,i} + n_{l,i}$ and the prediction block of $X_h$ is $\textbf{X}_l + \textbf{N}_l$.
The prediction error of $\textbf{X}_h$ with a zero MV is expressed as,
\begin{equation}
\label{equ:19}
pe = \sum_{x_{h,i}\in \textbf{X}_h} (x_{h,i}- x_{l,i}-n_{l,i})^2,
\end{equation}
where $pe$ denotes the prediction error.
It can be rewritten as,
\begin{equation}
\label{equ:20}
pe = \sum_{x_{h,i}\in \textbf{X}_h} (x_{h,i}- x_{l,i})^2 + n_{l,i}^2 - 2 n_{l,i} (x_{h,i}- x_{l,i}).
\end{equation}

Suppose $n_{l,i}$ and $(x_{h,i}- x_{l,i})$ are independent, we can have
\begin{equation}
\label{equ:21}
pe = M\{ E [(x_{h,i}- x_{l,i})^2]+ E[n_{l,i}^2] + 2 E [n_{l,i}] E [(x_{h,i}- x_{l,i})] \},
\end{equation}
where $E[\cdot]$ is the notation of the expectation, and $M$ denotes the number of samples in the CU $\textbf{X}_h$.
Since the coding errors can be assumed to have zero-mean \cite{zeromean}, i.e., $E [n_{l,i}]= 0$, we can obtain,
\begin{equation}
\label{equ:22}
pe = M\{ E [(x_{h,i}- x_{l,i})^2]+ E[n_{l,i}^2] \}.
\end{equation}

\subsection{Refined Prediction Error}

The far layers store the highest depth values and the corresponding near layers store the lowest depth values.
We would like to consider adding an integer offset $\delta$ to each pixel of the prediction block. That is, the prediction block is refined as $\textbf{X}_l + \textbf{N}_l + \textbf{C}$, where $\textbf{C}$ is a matrix in which all the elements are $\delta$.
The prediction error becomes,
\begin{equation}
\label{equ:23}
pe' = M\{ E [(x_{h,i}- x_{l,i}-\delta)^2]+ E[n_{l,i}^2] \},
\end{equation}
where $pe'$ denotes the refined prediction error.

The difference between $pe$ and $pe'$ is,
\begin{equation}
\label{equ:24}
pe - pe' =   2 M\delta \cdot E [ (x_{h,i}- x_{l,i})] - M \delta^2.
\end{equation}
We denote $\alpha, 0 \leq \alpha \leq 1$ as the proportion of the samples in $\textbf{X}_h$ which have the same depth values as $\textbf{X}_l$. That is, $\alpha M$ samples have the same depth value, i.e., $x_{h,i}= x_{l,i}$. Moreover, $(1-\alpha)M$ samples have different depth values, i.e., $x_{h,i} > x_{l,i}$. Since depth values are positive integers, the inequation $x_{h,i} > x_{l,i}$ is equivalent to $x_{h,i} - x_{l,i} \geq 1$. Then, we can obtain $E [(x_{h,i}- x_{l,i})]\geq (1-\alpha)$.
\begin{equation}
\label{equ:25}
pe - pe' \geq    M\delta (2-2\alpha-\delta).
\end{equation}

Since $M > 0$ and $0 \leq \alpha \leq 1$, to make $pe - pe' > 0$, the only feasible solution is $\delta = 1$ and $\alpha < \frac{1}{2}$.
That is, if less than half of the samples have the same depth values and adding 1 to all the prediction pixels, the refined prediction error is less than the original prediction error.
In contrast, if $\alpha > \frac{1}{2}$, the sign of the difference between the two prediction errors is undetermined.

\subsection{Occupancy Map-based Merge Prediction}

From the analysis above, if $\alpha > \frac{1}{2}$, the refined prediction error is not necessarily smaller than the original one. However, since more than half of the samples are with the same depth values, the prediction error is significantly small.
In this case, encoding the CUs with the skip mode, in which the residues are not encoded, generally can achieve high performance.

On the other hand, if $\alpha < \frac{1}{2}$, i.e., more than half of the samples have different depth values.
In this case, the prediction residues should be encoded to reduce the distortion, i.e., the merge mode or the other inter prediction modes should be used.
The analysis above shows that, if $\alpha < \frac{1}{2}$, the prediction error can be reduced by adding 1 to each pixel of the block.
Therefore, an improved merge prediction approach can be proposed to reduce the prediction error.

To be more specific, an occupancy-map-based (OM-based) merge prediction approach is proposed in this work.
In the OM-based merge prediction method, the prediction blocks are refined based on the occupancy map, i.e., not adding 1 to all the pixels.
The reason is that, in the generated geometry videos, the empty samples are generated by copying either the last column or the last row of the previous blocks. We call them unoccupied pixels. Moreover, the unoccupied pixels in the corresponding far and near layers are padding with the same depth values, i.e., $x_{h,i} = x_{l,i}$.
Therefore, it is unnecessary to change the prediction pixels of the unoccupied pixels. That is, in the proposed OM-based merge prediction method, only the prediction pixels of the non-empty samples are needed to be refined.

Note that, if the OM-based merge prediction method is adopted, an occupancy map is necessary for decoding the geometry video.
Although the occupancy map is available in the decoder, we still provide an alternative scheme that is independent of the occupancy map, denoted as \emph{non-OM-based merge} prediction.
In the non-OM-based merge method, the prediction blocks are changed by directly adding 1 to all the pixels.
That is, the occupancy map is unnecessary for the non-OM-based merge prediction method.

\section{Experimental Results}
\label{sec:experiment}

In this section, we first outline the detailed test conditions.
Then, we present the overall performance of the proposed AGSC scheme.
Since the proposed method consists of two modules, including the EPM-RDO method and the OM-based merge prediction.
Ablation experiments are provided to verify the efficiency of each module.

\subsection{Test Conditions}

The proposed method was implemented by updating the V-PCC reference software TMC2-RD4.0\cite{TMuC2} and the HEVC reference software HM16.18-SCM8.7 \cite{HM16}.
The executable files are provided at the link\footnote{https://github.com/jxiong-njupt/AGSC.git}.
The experiments were tested on a PC with an Intel (R) 2.40 GHz processor, 16 Gb RAM.
The test conditions were set to lossy geometry, lossy attribute, and random access, which are defined in V-PCC common test condition (CTC) \cite{schwarz2018ctc}.
The non-normative 3D MVP \cite{Lili2020TIP} was set to true.
We have tested five point clouds, including \emph{Soldier}, \emph{Queen}, \emph{Longdress}, \emph{Redandblack}, and \emph{Loot}.
Please note that, for each DPC, the first 32 frames were tested as a good representation of the whole sequences. Five pairs of quantization parameters \cite{schwarz2018ctc} were tested to cover a broad range of bitrates.

The proposed method is compared with the V-PCC anchor in terms of the Bj\o{}ntegaard Delta (BD) bitrate (BD-rate)\cite{BDrate}.
The PSNR of Luma, Cb, and Cr are used as the objective distortion metrics to evaluate the attribute information, while the PSNRs of point-to-plane error (D2) and point-to-point error (D1) are used to evaluate the geometry information.

\begin{figure*}[tbp]
  \centering
  \subfigure[]{
    \label{fig:subfig:1d}
    \includegraphics[width=4cm]{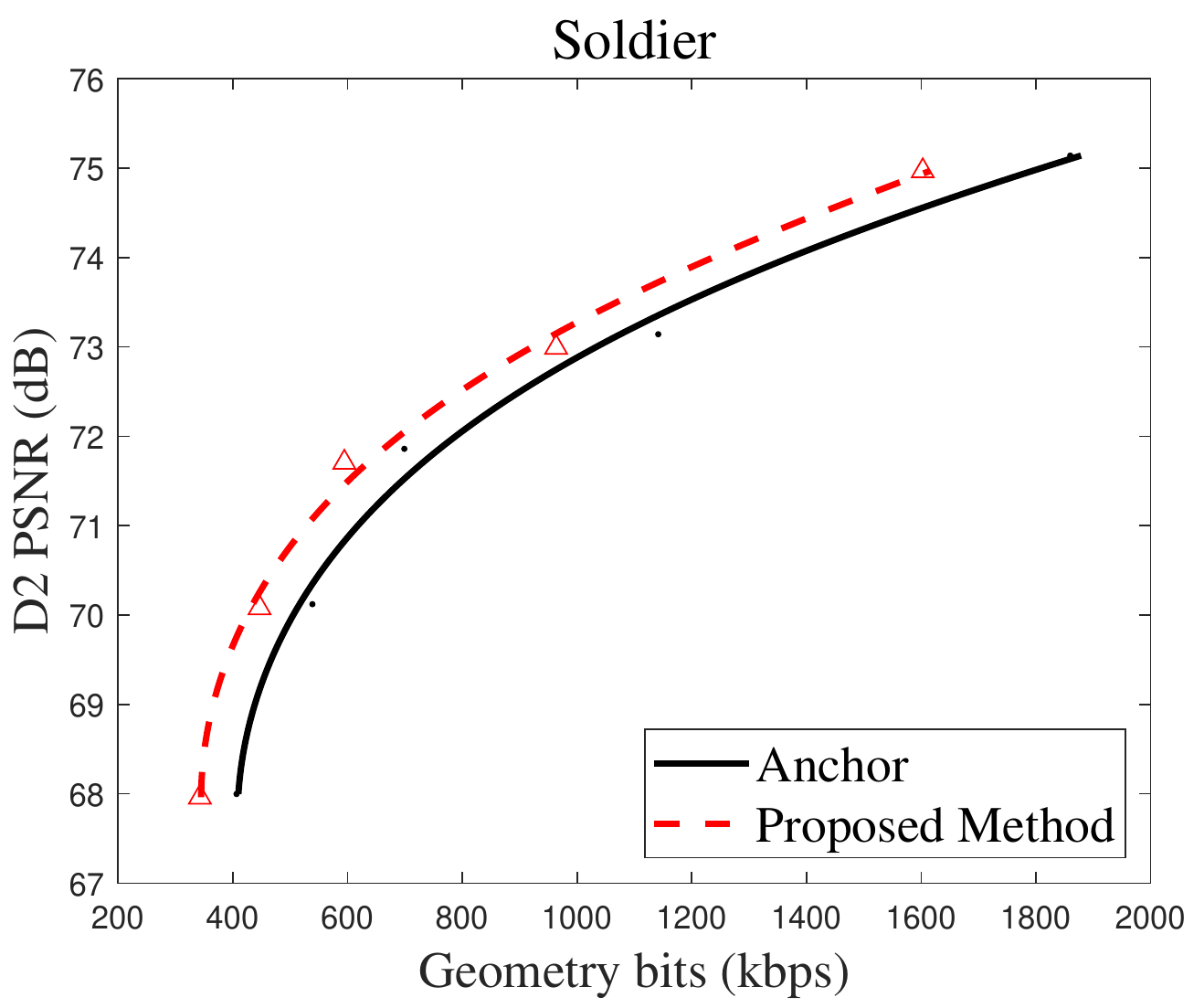}}
  \subfigure[]{
    \label{fig:subfig:1d}
    \includegraphics[width=4cm]{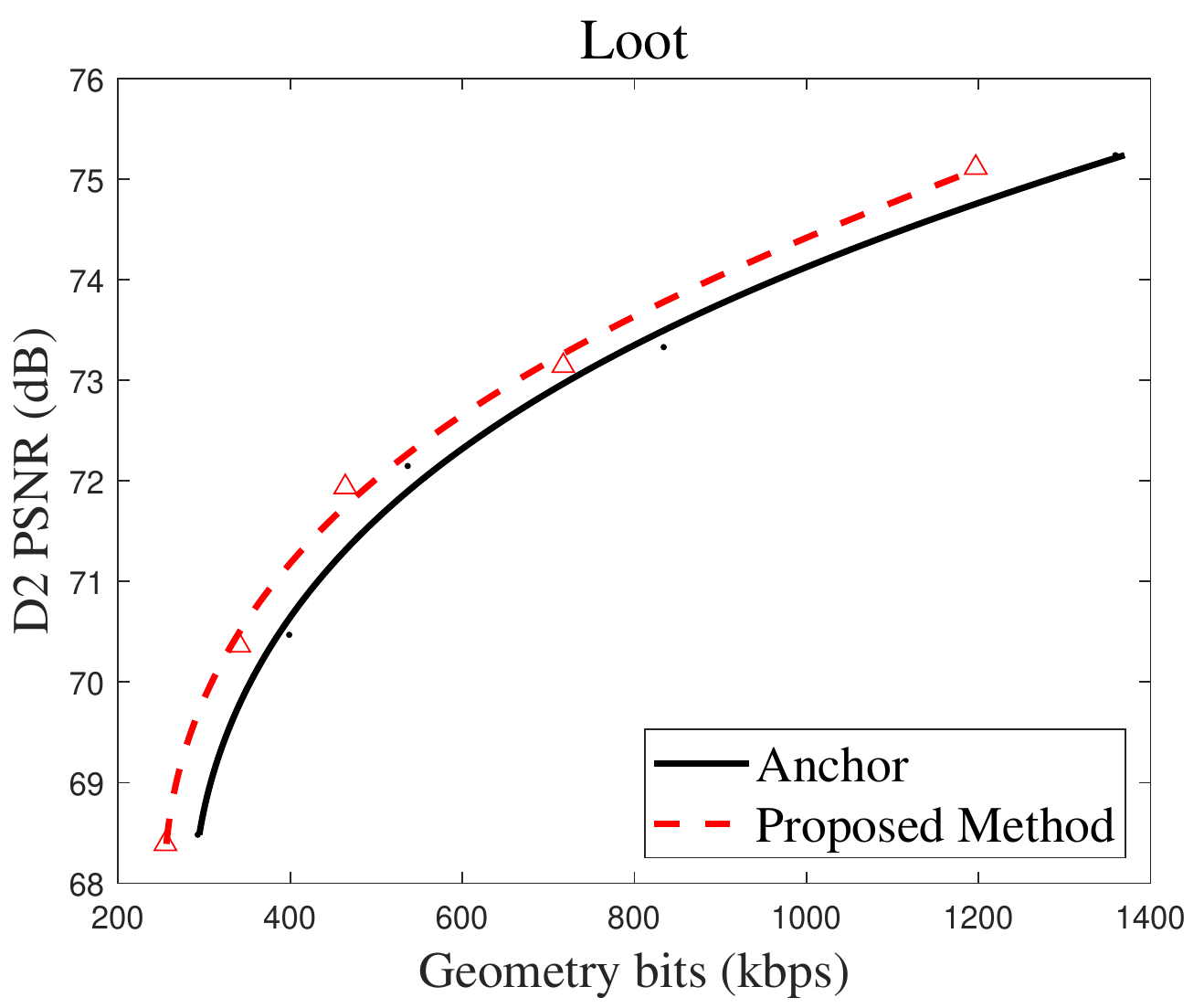}}
      \subfigure[]{
    \label{fig:subfig:1d}
    \includegraphics[width=4cm]{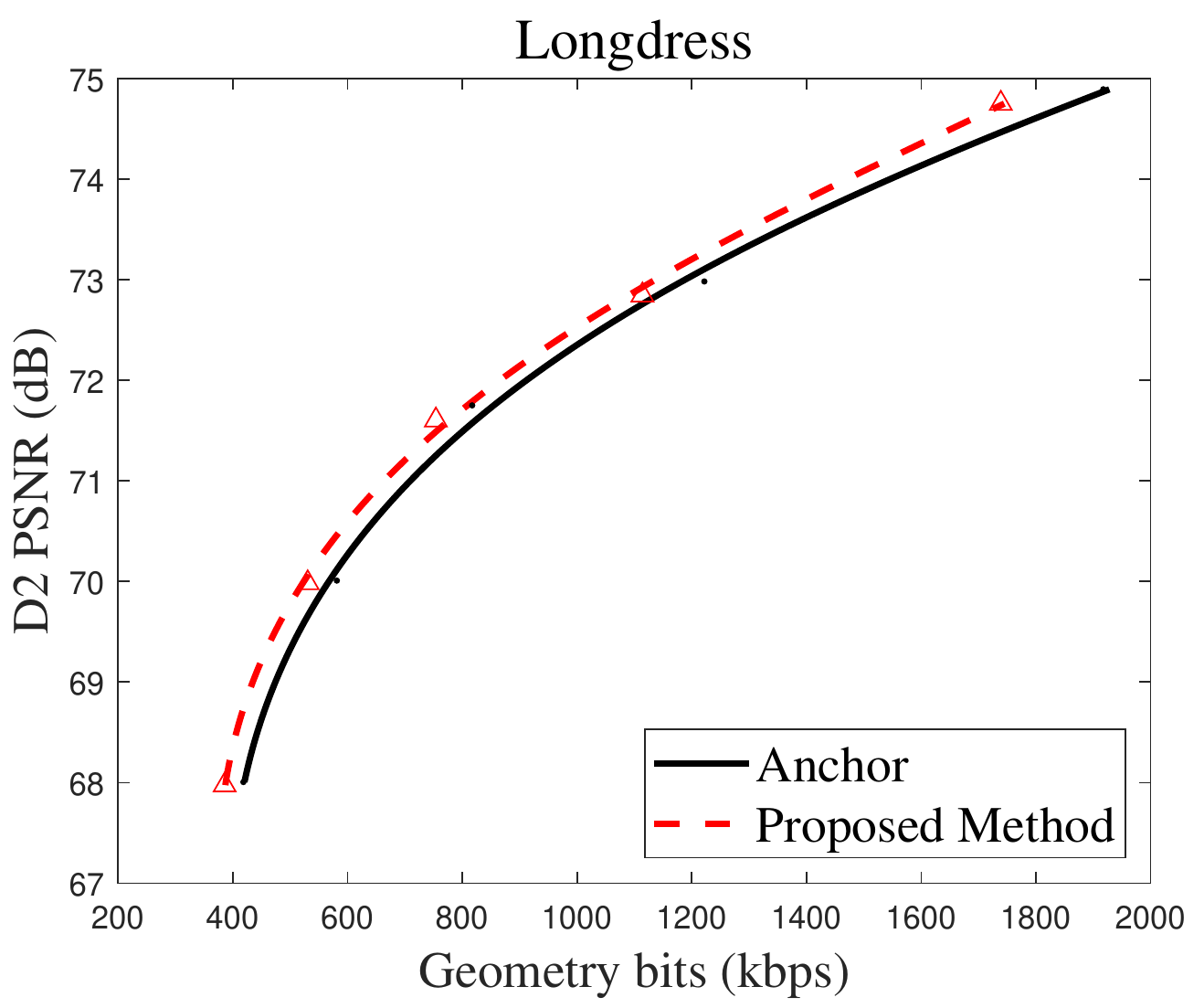}}
  \subfigure[]{
    \label{fig:subfig:1d}
    \includegraphics[width=4cm]{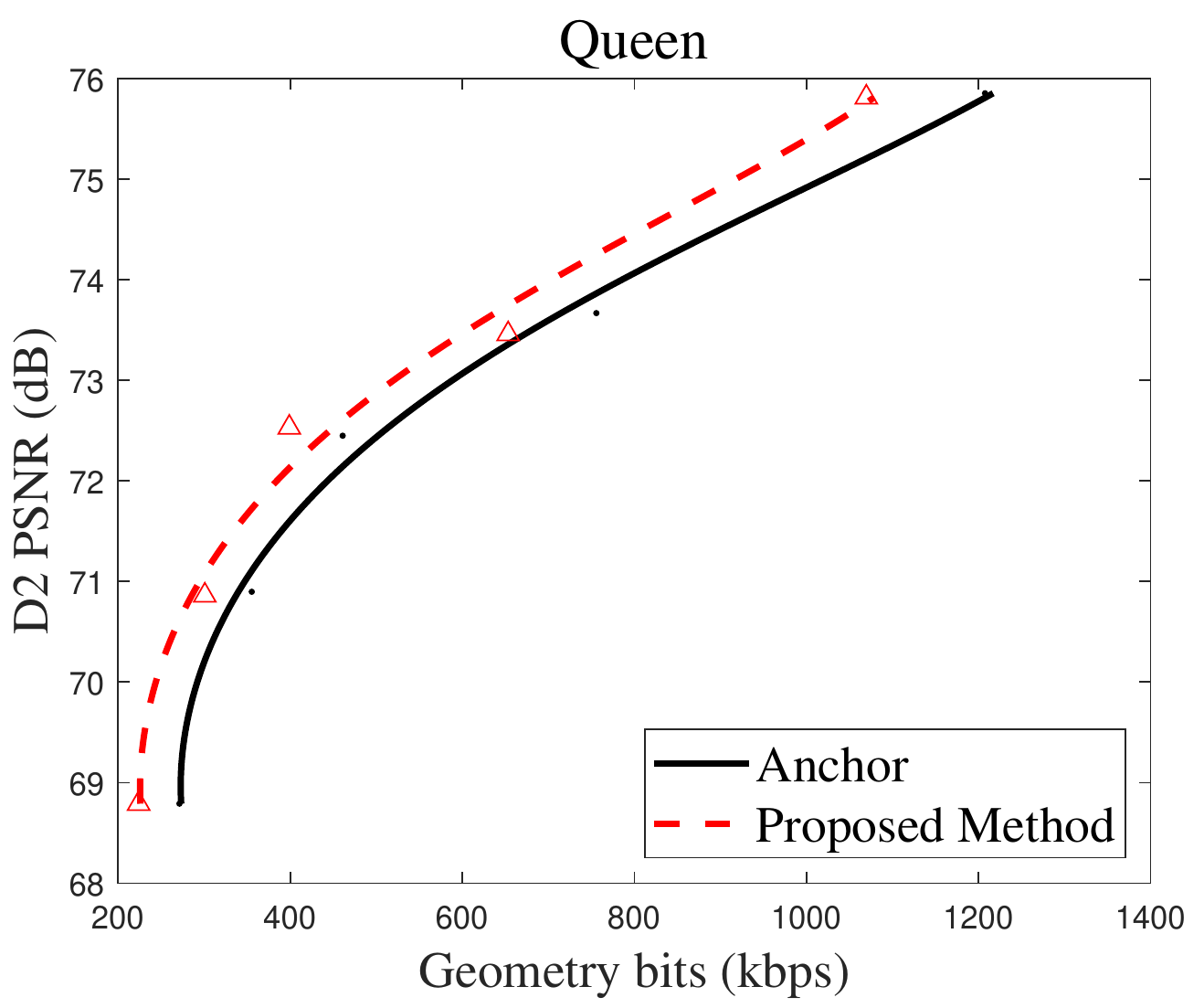}}
%
  \subfigure[]{
    \label{fig:subfig:1a}
    \includegraphics[width=4cm]{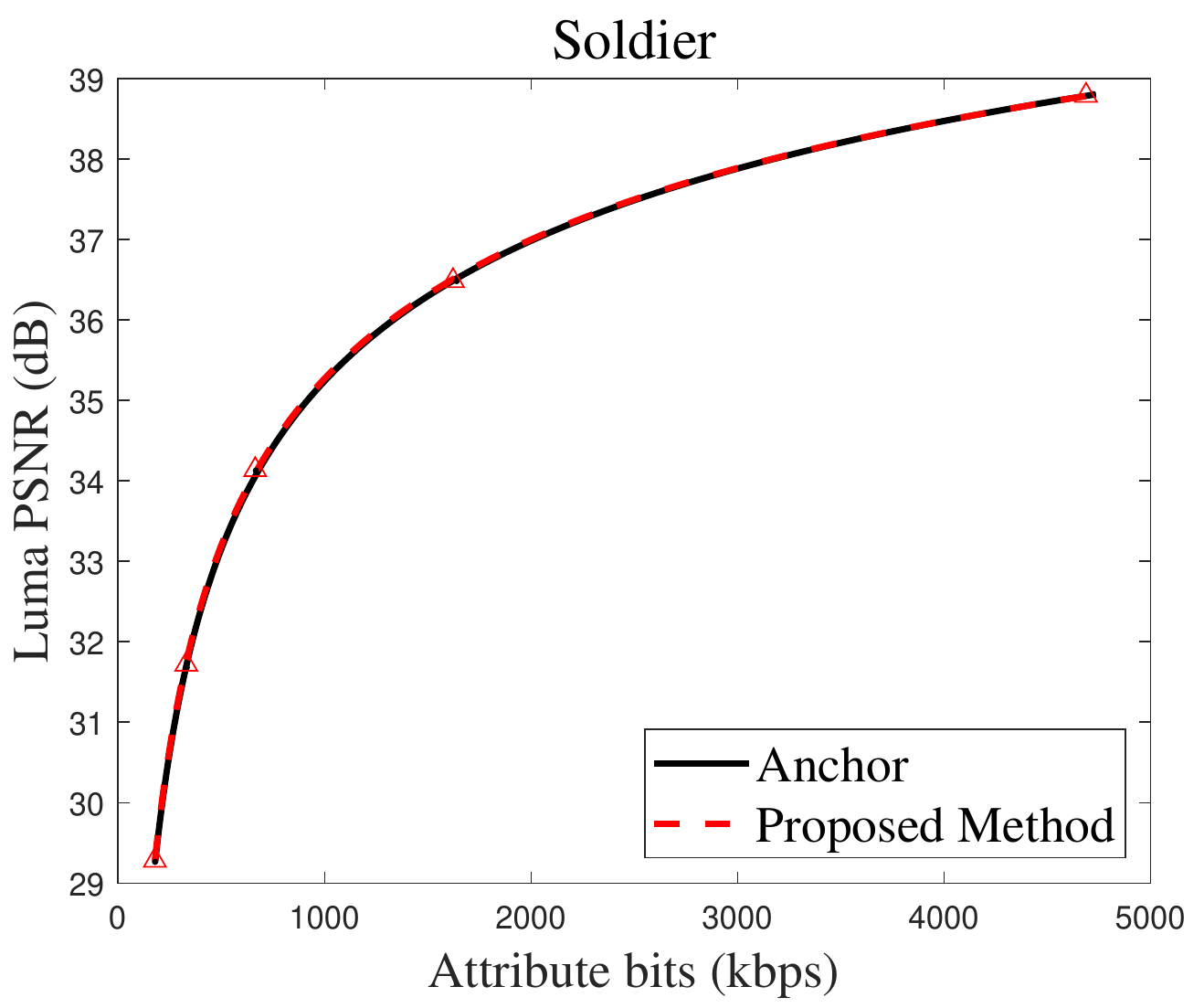}}
  \subfigure[]{
    \label{fig:subfig:1b}
    \includegraphics[width=4cm]{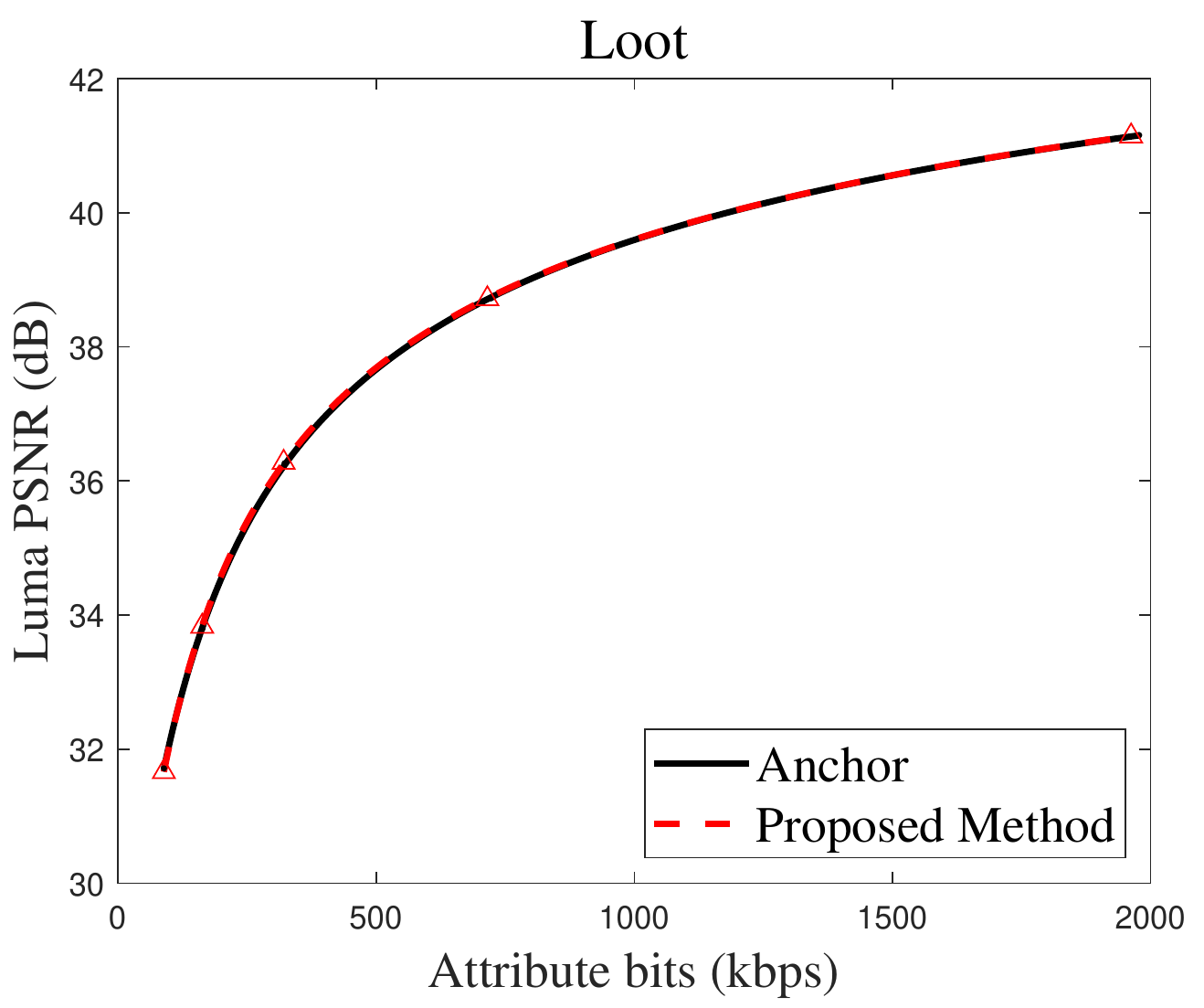}}
  \subfigure[]{
    \label{fig:subfig:1c}
    \includegraphics[width=4cm]{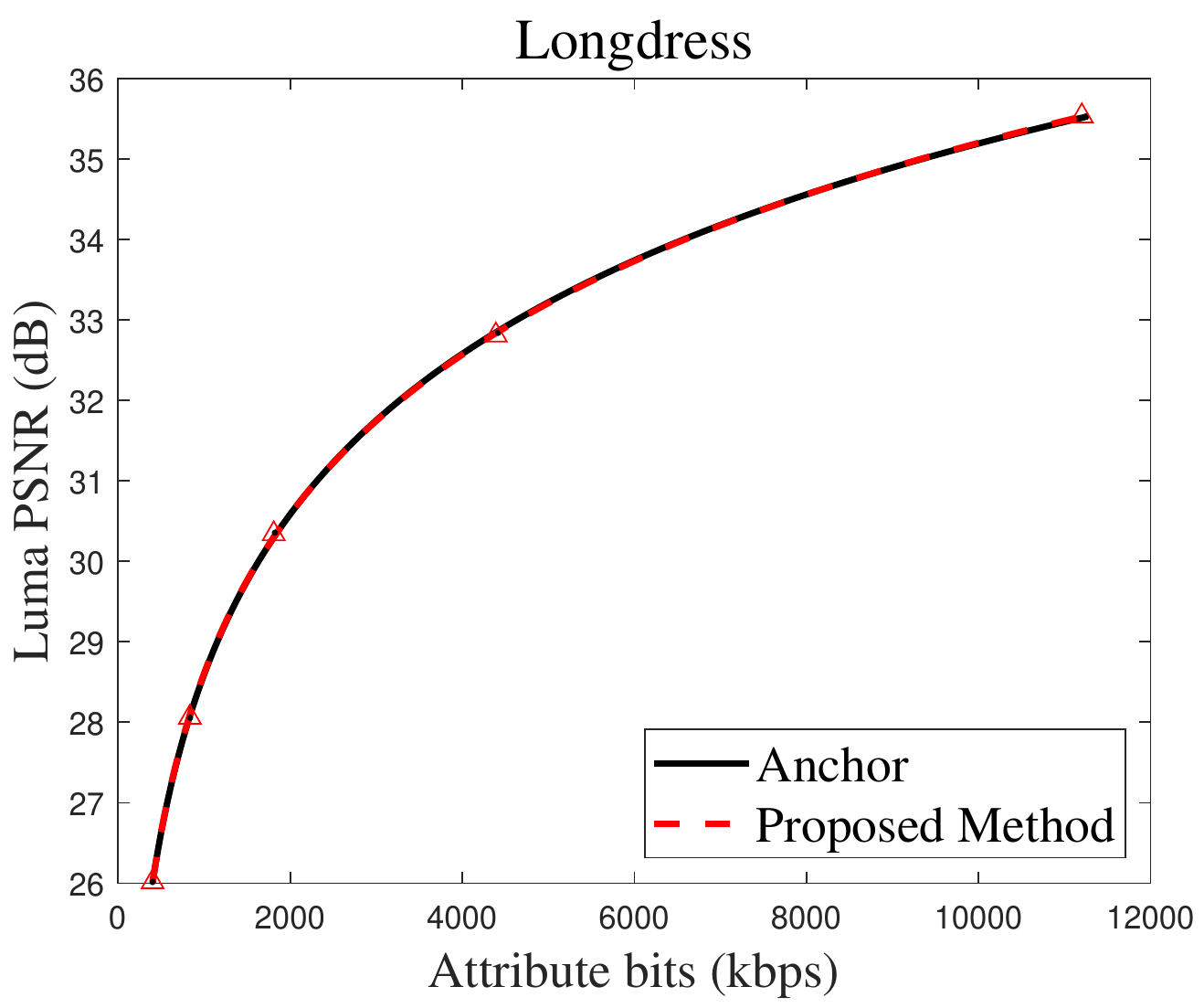}}
  \subfigure[]{
    \label{fig:subfig:1d}
    \includegraphics[width=4cm]{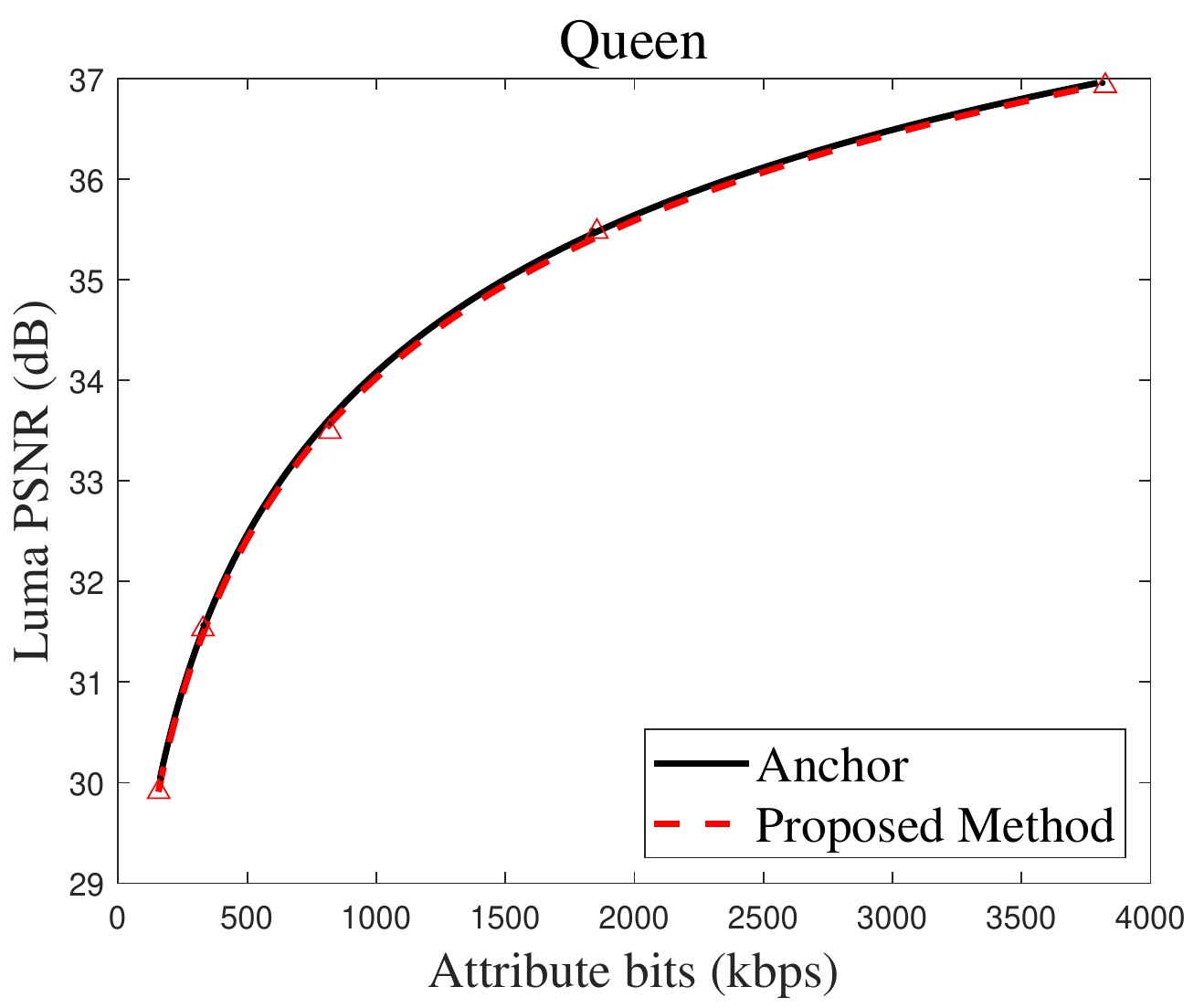}}
\caption{R-D curves of the proposed advanced geometry surface coding method compared with the V-PCC anchor.}
\label{fig:rdc}
\end{figure*}

\begin{table}[tbp]\centering											
\caption{PERFORMANCE OF THE PROPOSED ADVANCED GEOMETRY SURFACE CODING (AGSC) SCHEME COMPARED WITH THE V-PCC ANCHOR.}											
\scalebox{1}[1]{%
\begin{tabular}{c|cc|ccc}											
\toprule[1pt]											
 \ba{Tested Point Cloud}         &    \multicolumn{2}{c|}{Geom.BD-GeomRate}        & \multicolumn{3}{c}{Attr.BD-AttrRate}                         \\											
        &    D1    &    D2    &    Luma    &    Cb    &    Cr     \\ \hline											
Queen	& 	-13.40	& 	-13.22	& 	2.28	& 	0.28	& 	-1.60	\\
Loot	& 	-8.98	& 	-10.36	& 	-0.63	& 	-0.81	& 	-0.23	\\
Longdress	& 	-3.88	& 	-6.27	& 	-0.09	& 	-0.96	& 	-0.58	\\
Redandblack	& 	-3.96	& 	-5.99	& 	-0.42	& 	0.04	& 	-0.55	\\
Soldier	& 	-11.47	& 	-13.35	& 	-1.05	& 	-1.22	& 	-2.62	\\ \hline
Average	& 	-8.34	& 	-9.84	& 	0.02	& 	-0.53	& 	-1.11	\\ \hline
											
Enc. time self        &        \multicolumn{5}{c}{100\%}                                                                                \\											
Dec. time self        &        \multicolumn{5}{c}{100\%}                                                                                \\											
Enc. time children        &        \multicolumn{5}{c}{101.6\%}                                                                        \\											
Dec. time children        &        \multicolumn{5}{c}{100\%}                                                                            \\											
\bottomrule[1pt]											
\end{tabular}}											
\label{tab:OM-EPM}											
\end{table}			

\subsection{Overall Performance}

We first provide the overall performance of the proposed AGSC method.
The proposed method consists of two modules, including the EPM-based RDO and the OM-based merge prediction.
Table \ref{tab:OM-EPM} shows that the results of the proposed AGSC method compared with the V-PCC anchor.
It can be observed that the average bitrate savings of geometry information are 9.84\% and 8.34\% for the D2 and D1 metrics, respectively.
Whereas the bitrate of attribute information has a negligible change (increases 0.02\% on average) for the Luma component.
The attribute information even has some bitrate savings (0.53\% and 1.11\% on average) for the Cb and Cr components, respectively.
In particular, for the point cloud $\emph{Soldier}$, the geometry bitrate savings could be up to 13.35\% and 11.47\% for D2 and D1 metrics, respectively. Meanwhile, the attribute bitrate savings are 1.05\%, 1.22\%, and 2.62\%, for the Luma, Cb, and Cr components, respectively.
It can be concluded that reducing the geometry bitrate has no significant impact on the compression performance of attribute.
That is, the proposed AGSC method can significantly improve the geometry compression.

Fig. \ref{fig:rdc} shows the R-D curves of compressing the geometry information and attribute information.
The sub-figures Fig. \ref{fig:rdc} (a)-(d) show the R-D curves of compressing the geometry information.
The significant performance improvement for either the low bitrates or the high bitrates can be observed in these sub-figures.
The sub-figures Fig. \ref{fig:rdc} (e)-(h) show the R-D curves of compressing the attribute information.
It can be seen that the curves are completely overlapped.
That is, the proposed method has no significant impact on the compression of attribute information.
It can be concluded that the proposed AGSC method significantly improves the geometry compression without significant impact on the attribute compression.

We also compared the encoding time with the V-PCC anchor. As shown in Table \ref{tab:OM-EPM}, the Enc. time self and Dec. time self denote the encoding and decoding time of the V-PCC software.
Since there is no change in the V-PCC software, the time of V-PCC is not changed.
The Eec. time child and Dec. time child denote the coding and decoding time of the HEVC software.
It is observed that the encoding time is 101.6\% of the anchor time. That is for each CU, we only need to calculate the gradient and refine the prediction pixels of the merge modes.
Thus, the extra computation is negligible.

\subsection{Performance of EPM-based RDO}

\begin{table}[tbp]\centering											
\caption{PERFORMANCE OF THE PROPOSED EPM-BASED RDO METHOD COMPARED WITH THE V-PCC ANCHOR.}											
\scalebox{1}[1]{%
\begin{tabular}{c|cc|ccc}											
\toprule[1pt]											
 \ba{Tested Point Cloud}         &    \multicolumn{2}{c|}{Geom.BD-GeomRate}        & \multicolumn{3}{c}{Attr.BD-AttrRate}                         \\											
        &    D1    &    D2    &    Luma    &    Cb    &    Cr     \\ \hline											
Queen	& 	-7.60	& 	-7.35	& 	1.37	& 	1.66	& 	1.18	\\
Loot	& 	-2.20	& 	-6.81	& 	-0.30	& 	2.12	& 	-1.79	\\
Longdress	& 	0.23	& 	-4.09	& 	-0.35	& 	-0.47	& 	-0.50	\\
Redandblack	& 	0.12	& 	-4.12	& 	-0.44	& 	-0.24	& 	-0.61	\\
Soldier	& 	-5.20	& 	-8.59	& 	-0.50	& 	1.75	& 	-1.93	\\ \hline
Average	& 	-2.93	& 	-6.19	& 	-0.04	& 	0.96	& 	-0.73	\\ \hline
											
Enc. time self        &        \multicolumn{5}{c}{100\%}                                                                                \\											
Dec. time self        &        \multicolumn{5}{c}{100\%}                                                                                \\											
Enc. time children        &        \multicolumn{5}{c}{100.5\%}                                                                        \\											
Dec. time children        &        \multicolumn{5}{c}{100\%}                                                                            \\											
\bottomrule[1pt]											
\end{tabular}}											
\label{tab:EPM-RDO}											
\end{table}		

Table \ref{tab:EPM-RDO} shows the performance of the proposed EPM-based RDO method compared with the V-PCC anchor.
The average bitrate savings are 6.19\% and 2.93\% for D2 and D1 quality assessment metrics, respectively.
While the average bitrate savings are 0.04\%, -0.96\%, and 0.73\% for Luma, Cb, and Cr components, respectively.
In particular, the bitrate savings of the point cloud $\emph{Soldier}$ achieve at 8.59\% and 5.20\% for D2 and D1 metrics, respectively, and the bitrate saving of the Luma component is 0.50\%.
Thus, the proposed EPM-based RDO method can improve the coding performance of geometry information and has no significant impact on the coding of attribute information.

The results show that, for all the point clouds, the performance gains evaluated by the D2 metric are larger than those of the D1 metric. For instance, for the point cloud $\emph{Loot}$, the bitrate saving for the D2 metric is 6.81\% whereas the bitrate saving for the D1 metric is 2.20\%.
These results are reasonable because the proposed EPM model mainly focuses on the D2 error, which is the distance between the foot point and the evaluated point.
However, the D1 error is the distance between the corresponding point and the evaluated point.
As analyzed in section \ref{sec:ana}, the foot point is not necessarily the corresponding point.
However, for the point clouds $\emph{Queen}$ and $\emph{Soldier}$, there are significant performance gains for evaluating with both the two metrics.
This is because, as the nearest neighbor in the reference point cloud, the corresponding point is close to the foot point, i.e., the D1 and D2 errors are close to each other.

			
\subsection{Performance of Improved Merge Prediction Approaches}

Table \ref{tab:OM} shows the performance of the proposed OM-based merge prediction.
The average bitrate savings of geometry information are 4.61\% and 3.34\% for the D2 and D1 metrics, respectively.
Meanwhile, the average bitrate of the attribute information is slightly reduced (as 0.10\% for the Luma component).
That is, the results indicate the efficiency of the proposed OM-based merge prediction.

Besides the OM-based merge prediction, we also provide an alternative scheme, as non-OM-based merge prediction.
The results are shown in Table \ref{tab:non-OM}.
It demonstrates that the average bitrate savings of geometry information are 1.01\% and 0.23\% for the D1 and D2 metrics, respectively. Meanwhile, there is a slight bitrate saving of the attribute information.
That is, the performance gains of the non-OM based scheme are smaller than those of the OM-based scheme.
The results are reasonable. Because the OM-based scheme has accounted for the unoccupied pixels which have the same depth values in the corresponding far and near layers.
Whereas the non-OM-based scheme does not take this case into account.

To sum up, both the two modules of the proposed advanced geometry surface coding method, including the OM-based merge prediction and the EPM-based RDO, can significantly improve the compression performance. It is interesting that the overall performance gains of the proposed method (see Table \ref{tab:OM-EPM})) is greater than the sum of gains of the two modules (see Tables \ref{tab:EPM-RDO} and \ref{tab:OM}). That is, the two modules can be mutually reinforcing.

\begin{table}[tbp]\centering											
\caption{PERFORMANCE OF THE PROPOSED OM-BASED MERGE PREDICTION APPROACH COMPARED WITH THE V-PCC ANCHOR.}																	 \scalebox{1}[1]{%
\begin{tabular}{c|cc|ccc}											
\toprule[1pt]											
 \ba{Tested Point Cloud}         &    \multicolumn{2}{c|}{Geom.BD-GeomRate}        & \multicolumn{3}{c}{Attr.BD-AttrRate}                         \\											
        &    D1    &    D2    &    Luma    &    Cb    &    Cr     \\ \hline											
Queen	& 	-5.50	& 	-5.60	& 	0.41	& 	1.20	& 	-1.36	\\
Loot	& 	-5.35	& 	-3.64	& 	-0.24	& 	0.35	& 	-1.03	\\
Longdress	& 	-3.21	& 	-1.85	& 	-0.09	& 	-0.72	& 	-0.20	\\
Redandblack	& 	-3.79	& 	-1.70	& 	-0.29	& 	0.39	& 	-0.18	\\
Soldier	& 	-5.18	& 	-3.90	& 	-0.31	& 	1.09	& 	-1.17	\\ \hline
Average	& 	-4.61	& 	-3.34	& 	-0.10	& 	0.46	& 	-0.79	\\ \hline
											
Enc. time self        &        \multicolumn{5}{c}{100\%}                                                                                \\											
Dec. time self        &        \multicolumn{5}{c}{100\%}                                                                                \\											
Enc. time children        &        \multicolumn{5}{c}{101.2\%}                                                                        \\											
Dec. time children        &        \multicolumn{5}{c}{100\%}                                                                            \\											
\bottomrule[1pt]											
\end{tabular}}											
\label{tab:OM}											
\end{table}

\begin{table}[tbp]\centering											
\caption{PERFORMANCE OF THE PROPOSED NON-OM-BASED MERGE PREDICTION APPROACH COMPARED WITH THE V-PCC ANCHOR.}											
\scalebox{1}[1]{%
\begin{tabular}{c|cc|ccc}											
\toprule[1pt]											
 \ba{Tested Point Cloud}         &    \multicolumn{2}{c|}{Geom.BD-GeomRate}        & \multicolumn{3}{c}{Attr.BD-AttrRate}                         \\											
        &    D1    &    D2    &    Luma    &    Cb    &    Cr     \\ \hline											
Queen	& 	-1.07	& 	-1.53	& 	0.04	& 	-0.16	& 	-0.63	\\
Loot	& 	-1.04	& 	0.31	& 	-0.19	& 	-0.49	& 	-1.06	\\
Longdress	& 	-0.60	& 	0.25	& 	-0.24	& 	-0.46	& 	-0.39	\\
Redandblack	& 	-1.03	& 	0.41	& 	-0.26	& 	0.24	& 	-0.27	\\
Soldier	& 	-1.32	& 	-0.58	& 	-0.05	& 	-0.58	& 	-1.05	\\ \hline
Average	& 	-1.01	& 	-0.23	& 	-0.14	& 	-0.29	& 	-0.68	\\ \hline
											
Enc. time self        &        \multicolumn{5}{c}{100\%}                                                                                \\											
Dec. time self        &        \multicolumn{5}{c}{100\%}                                                                                \\											
Enc. time children        &        \multicolumn{5}{c}{101\%}                                                                        \\											
Dec. time children        &        \multicolumn{5}{c}{100\%}                                                                            \\											
\bottomrule[1pt]											
\end{tabular}}											
\label{tab:non-OM}											
\end{table}

\section{Conclusion}
\label{sec:conclusion}

In MPEG V-PCC, to compress the dynamic point clouds, the point clouds are decomposed into 2D videos to be compressed with the existing video codec, such as HEVC.
However, the existing video codec is originally designed for the natural visual signals, such that there are still problems in compressing the geometry information.
In this work, the advanced geometry surface coding method is proposed to solve the problems.
The proposed method consists of two modules, including the EPM-based RDO and the OM-based merge prediction.
Firstly, the 3D geometry information is evaluated with the point-to-point (D1) and point-to-plane (D2) error-based metrics, since the distortion model in existing RDO is not consistent with these metrics.
This paper presents an EPM model to describe the relationship between the existing distortion model and the D2 metric.
Then, we introduce an EPM-based RDO method in which the reconstruction error in the existing distortion model is projected on the plane normal. The proposed novel RDO is simplified to estimate the average normal vectors of CUs, such that it is feasible to implement the new RDO by weighting the existing distortion.
Secondly, to further reduce the prediction errors, we propose an OM-based merge prediction approach. In this method, the prediction pixels of occupied samples are refined by adding an offset based on the occupancy map.
Experimental results show that the proposed method can achieve an average 9.84\% bitrate saving for geometry compression.

\appendices

\ifCLASSOPTIONcaptionsoff

\fi
\bibliographystyle{IEEEtran}
\bibliography{reference}

\end{document}